\documentclass[preprint,showpacs,preprintnumbers,nofootinbib,amsmath,amssymb,aps,prc,longbibliography]{revtex4-1}
\usepackage{graphicx}% Include figure files
\usepackage{dcolumn}% Align table columns on decimal points
\usepackage{bm}% bold math
\begin{document}
%%%%%%%%%%%%%%%%%%%%%%%%%%%%%%%%%%%%%%%%%%%%%%%%%
\title{Thermodynamic properties of hot nuclei within \\ the self-consistent
quasiparticle random-phase approximation}
\author{N. Quang Hung$^{1}$}
 \email{nqhung@riken.jp}
 \author{N. Dinh Dang$^{2,3}$}
  \email{dang@riken.jp}
\affiliation{1) Center for Nuclear Physics, Institute of Physics, Hanoi,
Vietnam\\ 
2) Heavy-Ion Nuclear Physics Laboratory, RIKEN Nishina Center
for Accelerator-Based Science,
2-1 Hirosawa, Wako City, 351-0198 Saitama, Japan\\
3) Institute for Nuclear Science and Technique, Hanoi, Vietnam}
\date{\today}% It is always \today, today,
\begin{abstract}
%%%%%%%%%%%%%%%%%%%%%%%%%%%%%%%%%%%%%%%%%%%%%%%%%
    The thermodynamic properties of hot nuclei are described within the
    canonical and microcanonical ensemble approaches. These approaches are
    derived based on the solutions of the BCS and self-consistent
    quasiparticle random-phase approximation at zero temperature
    embedded into the canonical and microcanonical ensembles. The obtained
    results agree well with the recent data extracted from
    experimental level densities by Oslo group for $^{94}$Mo, $^{98}$Mo,
    $^{162}$Dy and $^{172}$Yb nuclei.
\end{abstract}

%Use showkeys class option if keyword
\maketitle

%%%%%%%%%%%%%%%%%%%%%%%%%%%%%%%%%%%%%%%%%%%%%%%%%
%%%%%%%%%%%%%%%%%%%%%%%%%%%%%%%%%%%%%%%%%%%%%%%%%%%%%%%
\section{INTRODUCTION}
\label{Intro}

Thermodynamic properties of highly excited (hot) nuclei have been a
topic of much interest in nuclear physics. From the theoretical
point of view, thermodynamic properties of any systems can be
studied by using three principal statistical ensembles, namely the
grand canonical ensemble (GCE), canonical ensemble (CE) and
microcanonical ensemble (MCE). The GCE is an ensemble of identical
systems in thermal equilibrium, which exchange their energies and
particle numbers with the external heat bath. In the CE, the systems
exchange only their energies, whereas their particle numbers are
kept to be the same for all systems. The MCE describes thermally
isolated systems with fixed energies and particle numbers. For
convenience, the GCE is often used in most of theoretical
approaches, e.g. the conventional finite-temperature BCS (FTBCS)
theory \cite{BCS}, and/or finite-temperature Hartree-Fock-Bogoliubov
theory \cite{HFB}. These theories, however, fail to describe
thermodynamic properties of finite small systems such as atomic
nuclei or ultra-small metallic grains. The reason is that the FTBCS
neglects the quantal and thermal fluctuations, which have been shown
to be very important in finite systems
\cite{Moretto,SPA,Zele,MBCS,FTBCS1,Ensemble}. These fluctuations
smooth out the superfluid-normal (SN) phase transition, which is a
typical feature of infinite systems as predicted by the FTBCS theory.

Because an atomic nucleus is a system with the fixed particle
number, the particle-number fluctuations are obviously not allowed. The use
of the GCE in nuclear systems is therefore an approximation, which is
good so long as the effect caused by
particle-number fluctuations are negligible. The
CE and MCE are often used in extending the exact solutions of the pairing
Hamiltonian \cite{Ensemble,Sumaryada,Exact} to finite temperature, whereas the CE is preferred
in the quantum Monte-Carlo calculations at finite temperature (FTQMC)
\cite{QMC,QMC1}. However, it is impracticable to find all
the exact eigenvalues of the pairing
Hamiltonian to construct the exact partition functions for
large systems. For instance, in the 
half-filled doubly-folded multilevel model (also called the
Richardson model) with $N=\Omega$ with $\Omega$
being the number of single-particle levels and $N$ - the number of
particles, this cannot be done already
for $N>$ 14 \cite{Ensemble,Sumaryada}.  Meanwhile, the FTQMC is quite time
consuming and cannot be applied to heavy nuclei unless a limited
configuration space is picked up. 
%%%% 3rd revision %%%%%%%
It is worth mentioning that the pairing Hamiltonian 
can also be solved exactly by using the Richardson's method, 
i.e. solving the Richardson equations. Using this method, 
the lowest eigenvalues of the pairing Hamiltonian can be obtained even 
for very large systems, e.g. with $N = \Omega$ = 1000
(See, e.g., Ref.~\cite{Dukelsky}).  Nonetheless, 
these lowest eigenstates (obtained after 
solving the Richardson equations) are not sufficient 
for the construction of the exact partition function
at finite temperature since the latter should contain all the excited
states, not only the lowest ones.
%%%%%%%%%%%%%%%%%%%
In principle, the CE-based approaches
can also be derived from the exact particle-number projection (PNP)
at finite temperature on top of the GCE ones \cite{FTPNP}. However, this
method is rather complicated for application to realistic nuclei. 

%%%% 2nd revision %%%%
The static path plus random phase approximation (SPA + RPA) 
with the exact number 
parity projection CSPA({\it p})~\cite{CSPA} and 
the latter extension of the number projected 
SPA (NPSPA)~\cite{NPSPA} offer quite good agreement 
with the exact CE of the Richardson model as well as the empirical heat capacities of 
heavy nuclei. However, Ref. \cite{CSPA} makes 
no comparison with experimental data, 
whereas Ref.~\cite{NPSPA} uses a thermal pairing gap, which is 
obtained from a direct extension of the odd-even mass difference to
finite temperature. As has been pointed in Ref. \cite{Ensemble} such
simple extension fails in the region of intermediate and high temperatures. 
In principle, the SPA can also be used to evaluate the MCE quantities 
based on the GCE ones by fixing the energy and particle number of the 
system~\cite{MCESPA}.  However this method is still quite 
complicated for practical applications to realistic nuclei, 
especially the heavy ones.
%%%%%%%%%%%%%%%%
From the experimental point of view, the CE and MCE are usually used to
extract various thermodynamic quantities of nuclear systems. This
is carried out by using the nuclear level density, which can be
experimentally measured at low excitation energy $E^{*}<$ 10 MeV.
Within the CE, the measured level densities are first extrapolated to
high $E^{*}$ using the back-shifted Fermi-gas model
(BSFG). The CE partition function is then constructed making use of
the Laplace transformation of the level density. Knowing the
partition function, one can calculate all the thermodynamic
quantities within the CE such as the free energy, total energy, heat
capacity and entropy. The thermodynamic quantities
of the systems obtained within the MCE are calculated via the Boltzmann's definition of
entropy. Although several experimental data for nuclear thermodynamic
quantities extracted in this way by the Oslo group have recently been
reported \cite{Oslo,Oslo1,Chankova,Kaneko}, most of present
theoretical approaches, derived within the GCE, cannot
describe well these data, which are extracted within the CE and MCE.
Recently we have proposed a method, which has allowed us to construct
theoretical approaches within the CE and MCE
to describe rather well thermodynamic
properties of atomic nuclei~\cite{CE-BCS}. The
proposed approaches are derived by solving the BCS
and self-consistent quasiparticle RPA (SCQRPA) equations with the
Lipkin-Nogami (LN) PNP for each total seniority $S$ (number of unpaired
particles at zero temperature)~\cite{SCQRPA}. The obtained results
are then embedded into the CE and MCE. Within the CE, the resulting
approaches are called the CE-LNBCS and CE-LNSCQRPA, whereas they are
called the MCE-LNBCS and MCE-LNSCQRPA within the MCE. The results
obtained within these approaches are found in quite good agreement with
not only the exact solutions of the Richardson model but
also the experimentally extracted data for $^{56}$Fe isotope. The
merit of these approaches reside in their simplicity and
feasibility in the application even to heavy
nuclei, where the exact solution is impracticable and the FTQMC is
time consuming. The goal of present article is to apply the
above-mentioned approaches to microscopically describe the recently
extracted thermodynamic quantities of $^{94, 96}$Mo, $^{162}$Dy and
$^{172}$Yb nuclei.

The article is organized as follows. The pairing Hamiltonian and the
derivations of the GCE-BCS, CE(MCE)-LNBCS and CE(MCE)-LNSCQRPA are
presented in {\rm Sec.} \ref{formalism}. The numerical results are
analyzed and discussed in {\rm Sec.} \ref{analysis}, whereas the
conclusions are drawn in the last section.

%%%%%%%%%%%%%%%%%%%%%%%%%%%%%%%%%%%%%%%%%%%%%%%%%
\section{FORMALISM}
%%%%%%%%%%%%%%%%%%%%%%%%%%%%%%%%%%%%%%%%%%%%%%%%%
\label{formalism}
\subsection{Pairing Hamiltonian}
\label{Hamil}
The present article considers the pairing Hamiltonian
\begin{equation}
H=\sum_{k\sigma=\pm}\epsilon_{k}a_{k\sigma}^{\dagger}a_{k\sigma}-
G\sum_{kk'}a_{k+}^{\dagger}a_{k-}^{\dagger}
a_{k'-}a_{k'+}~, \label{Hp}
\end{equation}
where $a_{k\sigma}^{\dagger}$ and $a_{k\sigma}$ are particle
creation and destruction operators on the $k$th orbitals,
respectively. The subscripts $k$ here imply the single-particle
states in deformed basis. This Hamiltonian describes a system of $N$
particles (protons or neutrons) interacting via a monopole pairing
force with constant parameter $G$.  The pairing Hamiltonian
\eqref{Hp} can be diagonalized exactly by using the SU(2) algebra of
angular momentum \cite{Exact}. At finite temperature $T \neq 0$, the
exact diagonalization is done for all total seniority or number of
unpaired particles $S$ because all excited states should be included
in the exact partition function. Here $S =$ 0, 2, $\ldots$ $N$ for
even-$N$ systems, and $S =$ 1, 3, $\ldots$ $N$-1 for odd-$N$
systems. For a system of $N$ particles moving in $\Omega$
degenerated single-particle levels, the number $n_{\rm Exact}$ of
exact eigenstates ${\cal E}_{i_S}^{\rm Exact}$ ($i_S =$1, \ldots,
$n_{\rm Exact}$) obtained within exact diagonalization is given as
\begin{equation}
n_{\rm Exact} = \sum_{S} {\rm C}_S^{\Omega} \times {\rm C}_{N_{\rm pair}-\frac{S}{2}}
^{\Omega - S} ~,
\label{nExact}
\end{equation}
which combinatorially increases  with $N$, where
$C_{n}^{m} = m!/[n!(m-n)!]$ and $N_{\rm pair}=N/2$ \cite{Ensemble}. Therefore,
the exact solution at $T \neq 0$ is impossible for large $N$
systems, e.g. $N >$ 14 for the half-filled case ($N=\Omega$), because
of the huge size of the matrix to be diagonalized. 
%%%%%%%%%%%%%%%%%%%%%%%%%%%%%%%%%%%%%%%%%%%%%%%%%
\subsection{\label{GCE-BCS} GCE-BCS}
%%%%%%%%%%%%%%%%%%%%%%%%%%%%%%%%%%%%%%%%%%%%%%%%%
The well-known finite-temperature BCS (FTBCS) approach to the
pairing Hamiltonian \eqref{Hp} is derived based on the variational
procedure, which minimizes the grand potential
\begin{equation}
\Omega=\langle{H}\rangle - T{\cal S} - \lambda N \hspace{5mm} \rm{so
\hspace{2mm} that} \hspace{5mm} \delta\Omega=0 ~, \label{Omega}
\end{equation}
where ${\cal S}$ is the entropy of the system at temperature $T$.
The chemical potential $\lambda$ is a Lagrangian
multiplier, which can be obtained from the equation that maintain
the expectation value of the particle-number operator to be equal to
the particle number $N$.
The expectation value $\langle{\cal O}\rangle$ denotes
the GCE average of the operator ${\cal O}$ \cite{MBCS} 
(the Boltzmann's constant $k_B$ is set to 1),
\begin{equation}
\langle{\cal O}\rangle \equiv \frac{{\rm Tr}[{\cal O}e^{-\beta
(H-\lambda N)}]}{{\rm Tr}e^{-\beta (H-\lambda N)}} ~, \hspace{5mm}
\beta=\frac{1}{T} ~. \label{O}
\end{equation}
The conventional FTBCS equations for the pairing gap $\Delta$ and
particle number $N$ are then given as
\begin{equation}
\Delta=G\sum_{k}u_{k}v_{k}(1-2n_{k})~, \hspace{5mm}
N=2\sum_{k}\left[(1-2n_{k})v_{k}^2+n_{k}\right] ~, \label{FTBCS}
\end{equation}
where the Bogoliubov's coefficients $u_{k}$, $v_{k}$, the quasiparticle
energy $E_{k}$ and the quasiparticle occupation number $n_{k}$ have the
usual form as
\begin{equation}
u_{k}^{2}=\frac{1}{2}\left(1+\frac{\epsilon_{k}-Gv_{k}^2-\lambda}{E_{k}}\right)~,
\hspace{5mm}
v_{k}^{2}=\frac{1}{2}\left(1-\frac{\epsilon_j-Gv_{k}^2-\lambda}{E_{k}}\right)
~. \nonumber
\end{equation}
\begin{equation}
E_{k} = \sqrt{(\epsilon_{k}-Gv_{k}^2-\lambda)^2+\Delta^2}~, \hspace{5mm}
n_{k} = \frac{1}{1+e^{\beta E_{k}}} ~. \label{uv}
\end{equation}
The systems of Eqs. \eqref{FTBCS} and \eqref{uv} are called the
GCE-BCS equations. The total energy, heat capacity and entropy
obtained within the GCE-BCS are given as
\begin{equation}
{\cal E} = 2\sum_{k}\left[(1-2n_{k})v_{k}^2+n_{k}\right] -
\frac{\Delta^2}{G} - G\sum_{k}(1-2n_{k})v_k^4 ~, \nonumber
\end{equation}
\begin{equation}
C = \frac{\partial{\cal E}}{\partial T}~, \hspace{5mm} {\cal S}= -2\sum_{k}\left[n_{k}{\rm ln}n_{k} + (1-n_{k}){\rm
ln}(1-n_{k})\right]~. \label{EGCE}
\end{equation}
%%%%%%%%%%%%%%%%%%%%%%%%%%%%%%%%%%%%%%%%%%%%%%%%%
\subsection{\label{CE-LNBCS} CE-LNBCS}
%%%%%%%%%%%%%%%%%%%%%%%%%%%%%%%%%%%%%%%%%%%%%%%%%
Different from the GCE-BCS, the CE-LNBCS is derived based on the
solutions of the BCS equations combined with the Lipkin-Nogami
particle-number projection (PNP) \cite{LN} at $T=0$ for each total
seniority $S$ of the system. When the pairs are broken, the unpaired
particles denoted with the quantum numbers $k_S$ block the
single-particle levels $k$. As the result, these blocked
single-particle levels do not contribute to the pairing correlation.
Therefore, the Lipkin-Nogami BCS (LNBCS) equations at $T=0$ can be
derived by excluding these $k_S$ blocked levels. These equations are given as
\begin{equation}
\Delta^{\rm LNBCS}(k_{S}) = G\sum_{k\neq k_S}u_{k}v_{k},
\hspace{5mm} N = 2\sum_{k\neq k_{s}}v_{k}^{2}+S ~, \label{Del}
\end{equation}
where
\begin{equation}
u_{k\neq k_S}^{2}= \frac{1}{2}\left(1+\frac{\epsilon_{k}-Gv_{k}^2-\lambda(k_{S})}{E_{k}}\right)~,
\hspace{5mm} v_{k\neq k_{S}}^2= \frac{1}{2}\left(1-\frac{\epsilon_{k}-Gv_{k}^2-\lambda(k_{S})}{E_{k}}\right)
~, \label{uvk}
\end{equation}
\begin{equation}
E_{k\neq k_S} = \sqrt{[\epsilon_{k}-Gv_{k}^2-\lambda(k_{S})]^2+[\Delta^{\rm
LNBCS}(k_{S})]^{2}} ~, \label{Ej}
\end{equation}
\begin{equation}
\lambda(k_{S}) = \lambda_{1}(k_{S}) + 2\lambda_{2}(k_{S})(N+1)~,
\hspace{5mm} \lambda_2(k_{S}) = \frac{G}{4}\frac{\sum_{k\neq
k_S}u_{k}^{3}v_{k} \sum_{k'\neq k'_S}u_{k'}v_{k'}^3 - \sum_{k\neq
k_S}u_{k}^{4}v_{k}^{4}}{(\sum_{k\neq k_S}u_k^2v_k^2)^2-\sum_{k\neq
k_S}u_k^4v_k^4} ~. \label{la}
\end{equation}
As for the blocked single-particle levels, $k=k_{S}$, their
occupation numbers are always equal to $1/2$. Solving the systems of
Eqs. \eqref{Del} - \eqref{la}, one obtains the pairing gap
$\Delta^{\rm LNBCS}(k_S)$, quasiparticle energies $E_k$ and
Bogoliubov's coefficients $u_k$ and $v_k$, which correspond to each
position of unpaired particles on blocked levels $k_S$ at each value
of the total seniority $S$. There are $n_{\rm LNBCS}=\sum_S
C_S^{\Omega}$ configurations of $k_S$ levels distributed amongst
$\Omega$ single-particle levels at each value of $S$, which is also
the number of eigenstates obtained within the LNBCS. The LNBCS
energy (eigenvalue) ${\cal E}_{i_{S}}^{\rm LNBCS}$ for each
configuration is then given as
\begin{equation}
{\cal E}_{i_{S}}^{\rm LNBCS} = 2\sum_{k\neq k_S}{\epsilon_k v_k^2} +
\sum_{k_S}\epsilon_{k_S} - \frac{[\Delta^{\rm LNBCS}(k_S)]^2}{G} -
G\sum_{k\neq k_S}v_k^4 - 4\lambda_2(k_S)\sum_{k\neq k_S}u_k^2v_k^2
~. \label{EBCS}
\end{equation}
The partition function of the so-called CE-LNBCS approach is
constructed by using the LNBCS eigenvalues ${\cal E}_{i_{S}}^{\rm LNBCS}$ as
\cite{CE-BCS}
\begin{equation}
Z_{\rm LNBCS}(\beta) = \sum_{S}d_{S}\sum_{i_{S}=1}^{n_{\rm LNBCS}}{e^{-\beta{\cal
E}_{i_{S}}^{\rm LNBCS}}}
~, \label{ZBCS}
\end{equation}
where $d_S = 2^S$ is the degeneracy. Knowing the partition function
\eqref{ZBCS}, we can calculate all thermodynamic quantities of the
system such as the free energy $F$, entropy ${\cal S}$, total energy
${\cal E}$, and heat capacity $C$ as follows
\begin{equation}
F = -T{\rm ln}Z(T), \hspace{5mm} {\cal S} = -\frac{\partial
F}{\partial T}~, \hspace{5mm} {\cal E} = F + T{\cal S}, \hspace{5mm}
C = \frac{\partial{\cal E}}{\partial T} ~. \label{FEC}
\end{equation}
The pairing gap is obtained by averaging the seniority-dependent gaps
$\Delta_{i_{S}}^{\rm LNBCS}=\Delta^{\rm LNBCS}(k_{S})$ at $T=0$ in the CE by means of the CE-LNBCS partition
function \eqref{ZBCS}, namely
\begin{equation}
\Delta_{\rm CE-LNBCS} = \frac{1}{Z_{\rm LNBCS}} \sum_{S}
d_S\sum_{i_{S}}^{n_{\rm LNBCS}}{\Delta^{\rm
LNBCS}_{i_{S}}e^{-\beta{\cal E}_{i_{S}}^{\rm LNBCS}}} ~.
\label{GapCEBCS}
\end{equation}

%%%%%%%%%%%%%%%%%%%%%%%%%%%%%%%%%%%%%%%%%%%%%%%%%
\subsection{\label{CE-QRPA} CE-LNSCQRPA}
%%%%%%%%%%%%%%%%%%%%%%%%%%%%%%%%%%%%%%%%%%%%%%%%%
As mentioned previously in {\rm sec.} \ref{Hamil}, a complete CE
partition function should include all eigenstates. The LNBCS theory
(at $T=0$) produces only the lowest eigenstates. For instance, for
even (odd) $N$ there is only one state at $S=$ 0, which is the
ground state. For $S>$ 0 there are also excited states in even
(odd) systems, whose total number n$_{\rm LNBCS}$ is much smaller
than n$_{\rm Exact}$. Consequently, the results obtained within the
CE-LNBCS can be compared with the exact ones only at low $T$ because
at high $T$, higher eigenstates (excited states), which the LNBCS
theory cannot reproduce, should be included in the CE partition
function. This can be done by going beyond the quasiparticle mean
field by introducing the SCQRPA with Lipkin-Nogami PNP (LNSCQRPA),
which incorporates not only the ground states but also the pairing
vibrational excited states predicted by the QRPA~\cite{SCQRPA}. The
derivation of the LNSCQRPA equations has been presented in details
in Refs. \cite{FTBCS1,SCQRPA,Chemical}, so we do not repeat it here.
The LNSCQRPA formalism at $T=0$ for each total seniority $S$ is
proceeded in the same way as that of the LNBCS described in the
previous section, namely the LNSCQRPA equations are derived only for
the unblocked levels $k\neq k_S$, whereas the levels, blocked by the
unpaired particles $k=k_S$, do not contribute to the pairing
Hamiltonian. The SCQRPA equations at $T=$ 0 has been derived in Ref.
\cite{SCQRPA}, whose final form reads
\begin{equation}\label{SCQRPA}
\left(\begin{array}{cc}A&B\\B&A\end{array}\right)
\left(\begin{array}{cc}X_k^\nu\\Y_k^\nu\end{array}\right)
=\omega_\nu\left(\begin{array}{cc}X_k^\nu\\-Y_k^\nu\end{array}\right)~,
\end{equation}
The SCQRPA submatrices are given as
%%%%%%%%%%%%%%%%%%%%%%%%%%%%%%%%%%%%%%%%%%%%%%%%
\[
A_{kk'} = 2\bigg[b_{k}+2q_{kk'}+2\sum_{k''}q_{kk''}
(1-{\cal D}_{k''})
    - \frac{1}{{\cal D}_{k}}
    \bigg(\sum_{k''}d_{kk''}
\langle\bar{0}|{\cal A}_{k''}^{\dagger}{\cal A}_{k}|\bar{0}\rangle
\]
\begin{equation}
    - 2\sum_{k''}h_{kk''}\langle\bar{0}|{\cal A}_{k''}
{\cal A}_k|\bar{0}\rangle\bigg)\bigg]\delta_{kk'}
+ d_{kk'}\sqrt{{\cal D}_{k}{\cal D}_{k'}}
+ 8q_{kk'}\frac{\langle\bar{0}|{\cal A}_{k}^{\dagger}{\cal A}_{k'}|\bar{0}\rangle}
{\sqrt{{\cal D}_{k}{\cal D}_{k'}}}~,
\label{A}
\end{equation}
\[
B_{kk' }= -2\bigg[h_{kk'}+
\frac{1}{{\cal D}_{k}}\bigg(\sum_{k''}d_{kk''}
\langle\bar{0}|{\cal A}_{k''}{\cal A}_{k}|\bar{0}\rangle
+ 2\sum_{k''}h_{kk''}
\langle\bar{0}|{\cal A}_{k''}
^{\dagger}{\cal A}_{k}|\bar{0}\rangle\bigg)\bigg]\delta_{kk'}
\]
\begin{equation}
    + 2h_{kk'}\sqrt{{\cal D}_{k}{\cal D}_{k'}}
+ 8q_{kk'}\frac{\langle\bar{0}|{\cal A}_{k}{\cal A}_{k'}
|\bar{0}\rangle}{\sqrt{{\cal D}_{k}{\cal D}_{k'}}}
\label{B}~,
\end{equation}
where $b_{k}$, $d_{kk''}$, $h_{kk''}$, and $q_{kk'}$ (all $k\neq
k_{S}$) are functions of
$u_{k}$, $v_{k}$, $\epsilon_{k}$, $\lambda$ and $G$ as given in Eqs.
(13), (15), (17) and (18) of Ref. \cite{SCQRPA}.
The screening factors
$\langle\bar{0}|\mathcal{A}_k^{\dagger}\mathcal{A}_{k'}|\bar{0}\rangle$ and
$\langle\bar{0}|\mathcal{A}_k\mathcal{A}_{k'}|\bar{0}\rangle$ with
${\cal A}^{\dagger}\equiv\alpha^{\dagger}_{k}\alpha^{\dagger}_{-k}$
being the creation operator of two-quasiparticle pair are given in terms of
the SCQRPA amplitudes $\mathcal{X}_k^{\nu}$ and
$\mathcal{Y}_k^{\nu}$ as
\begin{equation}
\langle\bar{0}|\mathcal{A}_k^{\dagger}\mathcal{A}_{k'}|\bar{0}\rangle
=\sqrt{\langle{\cal D}_{k}\rangle\langle{\cal D}_{k'}
\rangle}\sum_{\nu}{\mathcal{Y}_k^{\nu}\mathcal{Y}_{k'}^{\nu}}~,
\hspace{5mm}
\langle\bar{0}|\mathcal{A}_k\mathcal{A}_{k'}|\bar{0}\rangle
=\sqrt{ \langle{\cal D}_{k}\rangle
\langle{\cal
D}_{k'}\rangle}\sum_{\nu}{\mathcal{X}_k^{\nu}\mathcal{Y}_{k'}^{\nu}}~.
\end{equation}
where $\langle\bar{0}|\ldots|\bar{0}\rangle$ denotes the expectation
value in the SCQRPA ground state. The ground-state correlation factor
${\cal D}_{k}$ is expressed in term of the backward-going amplitudes ${\cal
Y}^{\nu}_{k}$ as ${\cal D}_{k}=[1+2\sum_{\nu}({\cal
Y}^{\nu}_{k})^{2}]^{-1}$ with the sum running over all the SCQRPA
solutions $\nu$.

After solving the LNSCQRPA equations (\ref{Del}), (\ref{SCQRPA}) --
(\ref{B}) for each total seniority $S$, we obtain a set of
eigenstates, which consists of C$^{\Omega}_S$ lowest eigenstates (the
ground state at $S=$0 or 1), as well as higher eigenstates
(excited states) on top of these lowest ones, which come from the
solutions of the LNSCQRPA equations, whose eigenvalues are
$\omega_{\nu}^{(S)}$ ($\nu=1,\ldots\Omega-S$)\footnote{The SCQRPA
has altogether $\Omega-S +1$ solutions with positive energies.
However the lowest one corresponds to the spurious mode, whose
energy is zero within the QRPA. Therefore it is excluded in the
numerical calculations.}. As the result, the total number of
eigenstates obtained within the LNSCQRPA is given as
\begin{equation}
n_{\rm LNSCQRPA} = \sum_S{\rm C}_S^{\Omega} \times (\Omega - S) ~.
\label{nLNSCQRPA}
\end{equation}
Consequently, the
so-called CE-LNSCQRPA partition function is calculated as
\begin{equation}
Z_{\rm LNSCQRPA}(\beta) = \sum_S{d_S}\sum_{i_{S}=1}^{n_{\rm
LNSCQRPA}} e^{-\beta{\cal E}_{i_{S}}^{\rm LNSCQRPA}} ~,
\label{ZQRPA}
\end{equation}
which is formally identical to the CE-LNBCS partition function
\eqref{ZBCS}, but the LNBCS eigenvalues ${\cal E}_{i_{S}}^{\rm
LNBCS}$ are now replaced with ${\cal E}_{i_{S}}^{\rm LNSCQRPA}$.
From this partition function, the thermodynamic quantities obtained
within the CE-LNSCQRPA are calculated in the same way as those in
Eq. \eqref{FEC}. Although the number $n_{\rm LNSCQRPA}$ of the
LNSCQRPA eigenstates is larger than $n_{\rm LNBCS}$, it is still
much smaller than $n_{\rm Exact}$. This most important feature of
the present method tremendously reduces the computing time in
numerical calculations for heavy nuclei. As an example, we show in
Table \ref{Table} the number of eigenstates and the total executing
time (the elapsed real time) for the exact diagonalization of the
pairing Hamiltonian, CE-LNBCS and CE-LNSCQRPA calculations 
within the Richardson model at several values $N$ of
particle number, which is taken to be equal to number $\Omega$ of
single-particle levels  (the half-filled case). This table shows
that the execution time within the LNSCQRPA (LNBCS) is shorter than
that consumed by the exact diagonalization by about two (four) orders.
%%%%%%%%%%%%%%%%%
\begin{table}[h]
\begin{center}
%%% 2nd revision %%%%%
    \caption{\label{Table}Number of eigenstates and
computation time for the exact diagonalization of the pairing Hamiltonian
as well as the numerical calculations within the CE-LNBCS and CE-LNSCQRPA
for the doubly-folded equidistant multilevel pairing
model at several values of  $N=\Omega$. The computation time is
estimated based on a shared large memory computer Altix 450 with
512GB memory of RIKEN Integrated Cluster of Clusters (RICC) system.}
\begin{tabular}{|c|ccc|cccc|}
\hline\hline
 && Number of eigenstates &&&~~~~~~Computation time&& \\\hline
$N$ & Exact & LNBCS & LNSCQRPA  &  ~~Exact & LNBCS & LNSCQRPA& \\\hline
$10$ & 8953 & 512 & 2560 &~~ 1 $ {\rm hr}$ & 1 ${\rm sec.}$ & 10 ${\rm sec.}$&\\
$12$ & 73789 & 2048 & 12288 &~~ 10 $ {\rm hrs}$ & 10 ${\rm sec.}$ & 1 ${\rm min.}$& \\
$14$ & 616227 & 8192 & 57344 &~~ 24 ${\rm hrs}$ & 1 ${\rm min.}$ & 10 ${\rm min.}$ &\\
$16$ & 5196627 & 32768 & 262144 & ~~- & 10 ${\rm min.}$ & 1 $ {\rm hr}$ &\\
$18$ & 44152809 & 131072 & 1179648 & ~~- & 1 $ {\rm hr}$ & 3 $ {\rm hrs}$ &\\
$20$ & 377379369 & 524288 & 5242880 & ~~- & 3 $ {\rm hrs}$ & 10 $ {\rm hrs}$& \\
\hline\hline
\end{tabular}

\end{center}
\end{table}

%%%%%%%%%%%%%%%%%%%%%%%%%%%%%%%%%%%%%%%%%%%%%%%%%
\subsection{\label{MCE} MCE-LNBCS, MCE-LNSCQRPA}

The MCE entropy is calculated by using the
Boltzmann's definition
\begin{equation}
{\cal S}({\cal E}) = {\rm ln} {\cal W}({\cal E}) ~,\hspace{5mm}
{\cal W}({\cal E}) = \rho({\cal E})\delta{\cal E}~, \label{SMCE}
\end{equation}
where $\rho({\cal E})$ is the density of states. In the LNBCS (LNSCQRPA) approaches,
${\cal W}({\cal E})$ is the number of LNBCS (LNSCQRPA)
eigenstates within the energy interval (${\cal E}, {\cal
E}+\delta{\cal E})$~\cite{Ensemble}. Knowing the MCE entropy,
one can calculate the MCE temperature as the first derivative of the
MCE entropy with respect to the excitation energy ${\cal E}$, namely
\begin{equation}
T = \left[\frac{\partial {\cal S}({\cal E})}{\partial{\cal
E}}\right]^{-1} ~. \label{TMCE}
\end{equation}
The corresponding approaches, which embed the LNBCS and LNSCQRPA
eigenvalues into the MCE, are called the MCE-LNBCS and MCE-LNSCQRPA,
respectively.
%%%%%%%%%%%%%%%%%%%%%%%%%%%%%%
\subsection{Level density}
\label{level}
The inverse relation of Eq. (\ref{SMCE}) reads
\begin{equation}
\rho({\cal E}) = e^{{\cal S}({\cal E})}/\delta{\cal E} ~,
\label{RhoMCE}
\end{equation}
which can be used to calculate the density of states $\rho({\cal E})$ from the fitted
MCE entropy.

Within the CE,
the density of states $\rho({\cal E})$ is calculated by using the method of
steepest descent to find the minimum of
the Laplace transform of the
partition function~\cite{Ericson}. As a result the density of states
$\rho({\cal E}$) at
temperature $T=\beta_{0}^{-1}$, which corresponds to this minimum, is approximated
as
\begin{equation}
\rho({\cal E}) \approx {Z(\beta_{0})e^{\beta_{0} {\cal
E}}}\bigg[2\pi\frac{\partial^{2}{\rm
ln}Z(\beta_{0})}{\partial\beta_{0}^{2}}\bigg]^{-1/2}
\equiv e^{{\cal S}(\cal E)}
\bigg(-2\pi\frac{\partial{\cal E}}{\partial\beta_{0}}~\bigg)^{-1/2} ~, \label{RhoCE}
\end{equation}
where $Z(\beta_{0})$, ${\cal S}({\cal E})$ and ${\cal E}$ are the CE partition function, entropy
and total excitation energy of the systems, respectively. The
density of states $\rho({\cal E})$ is obtained within the CE-LNBCS and CE-LNSCQRPA by replacing
the partition function $Z$ in Eq. \eqref{RhoCE} with that obtained
within the CE-LNBCS in Eq. \eqref{ZBCS} and CE-LNSCQRPA in Eq.
\eqref{ZQRPA}.

At finite angular momentum $J$, in principle, the approach of
LNSCQRPA plus angular momentum, which has been proposed by us in
Ref. \cite{SCQRPAM}, should be used to calculate the
angular-momentum dependent level density $\rho({\cal E},M)$ with $M$
being the $z$-projection of the total angular momentum. In this case
the former doubly-degenerated quasiparticle levels are resolved
under the constraint $M = \sum_{k}m_{k}(n_{k}^{+}-n_{k}^{-})$ with
the quasiparticle occupation numbers $n_{k}^{\pm}$, which are
described by the Fermi-Dirac distribution $n_{k}^{\pm,FD}=\{{\rm
exp} [\beta(E_{k} \mp \gamma m_{k})]+1\}^{-1}$ within the
non-interacting quasiparticle approximation, where $m_{k}$ is the
spin-projections of the $k$th single-particle state $|k,\pm
m_{k}\rangle$, $E_{k}$ is the quasiparticle energy, and $\gamma$ is
the rotation frequency. Knowing $\rho({\cal E},M)$, one can find
$\rho({\cal E},J) = \rho({\cal E}, M=J) - \rho({\cal E}, M = J+1)$
in the general case, where the total angular momentum $J$ is not
aligned with the $z$-axis~\cite{Bohr}. The total level density
$\rho_{tot}({\cal E})$ and experimentally observed level density
$\rho_{obs}({\cal E})$, are then defined as~\cite{Gilbert}
\begin{equation}
    \rho_{tot}({\cal E})=\sum_{J}(2J+1)\rho({\cal E},J)~,\hspace{5mm}
    \rho_{obs}({\cal E})=\sum_{J}\rho({\cal E},J)~.
    \label{rhoJ}
    \end{equation}
The empirical entropy ${\cal S}_{obs}({\cal E})$ is extracted from the observed level density
$\rho_{obs}({\cal E})$ in the same way as in Eq. (\ref{SMCE}),
replacing $\rho({\cal E})$ with $\rho_{obs}({\cal E})$, namely
\begin{equation}
{\cal S}_{obs}({\cal E}) = {\rm ln}[\rho_{obs}({\cal E})\delta{\cal
E}]~,
\label{Sobs}
\end{equation}
Because the
present article considers non-rotating nuclei at low angular momentum,
we assume that $\rho({\cal E},J)\simeq\rho({\cal E}, 0)\equiv\rho({\cal
E})$.
Therefore, by fitting the MCE entropy ${\cal S}({\cal E})$ in Eq.
(\ref{SMCE}) to the
experimentally observed entropy ${\cal S}_{obs}({\cal E})$ in Eq. (\ref{Sobs}),
i.e. ${\cal S}({\cal E})\simeq{\cal S}_{obs}({\cal E})$,
and inverting the obtained
result by using Eq. (\ref{RhoMCE}), what we get is actually the level density
comparable to the experimentally observed one,
$\rho_{obs}({\cal E})={\rm exp}[{\cal S}({\cal E})]/\delta{\cal E}$.
This means that the density of states $\rho ({\cal E})$ calculated by
using Eq. (\ref{RhoMCE}) or Eq. (\ref{RhoCE}) without taking into
account the effect of finite angular momentum is identical to the
level density like $\rho_{obs}({\cal E})$, not the total level density
$\rho_{tot}({\cal E})$, because of the absence of the factor $(2J+1)$.
%%%%%%%%%%%%%%%%%%%%%%%%%%%%%%%%%%%%%%%%%%%%%%%%%
\section{\label{analysis} ANALYSIS OF NUMERICAL RESULTS}
%%% 2nd revision %%%
The proposed approaches are used to calculate
the pairing gap, total energy, entropy and heat
capacity within the CE and MCE for a number of heavy isotopes, namely
$^{94, 98}$Mo, $^{162}$Dy and $^{172}$Yb \footnote{See, e.g. Fig. 1 of Ref. \cite{CE-BCS}
and Appendix A of the present article for the accuracy of the present approaches in comparison 
with the exact solutions of the Richardson model.}. The single-particle
energies are taken from the axially deformed Woods-Saxon potential
with the depth of the central potential \cite{WS}
\begin{equation}
V = V_0 \left[1 \pm k\frac{N-Z}{N+Z}\right] ~, \label{VWS}
\end{equation}
where $V_0=$ 51.0 MeV, $k=$ 0.86, whereas the plus and minus signs stand
for proton ($Z$) and neutron ($N$), respectively. The radius $r_0$,
diffuseness $a$, and spin-orbit strength $\lambda$ are chosen to be $r_0
=$ 1.27 fm, $a =$ 0.67 fm and $\lambda =$ 35.0. The quadrupole
deformation parameters $\beta_2$ are estimated from the experimental
$B(E2; 2^+_1 \rightarrow 0^+_1)$, which are 0.15, 0.17, 0.281 and
0.296 for $^{94}$Mo, $^{98}$Mo, $^{162}$Dy and $^{172}$Yb,
respectively \cite{Kaneko}. The pairing interaction parameters $G$
are adjusted so that the pairing gaps for neutrons and protons
obtained within the LNSCQRPA at $T=$ 0 and $S=$ 0 reproduce the values
extracted from the experimental odd-even mass differences,
namely $\Delta_N \simeq$ 1.2, 1.0,
0.8 and 0.8 MeV for neutrons, and $\Delta_Z \simeq$ 1.4,
1.3, 0.9 and 0.9 MeV for protons in $^{94}$Mo, $^{98}$Mo,
$^{162}$Dy and $^{172}$Yb, respectively.

It is well-known that pairing is significant only for the levels
around the Fermi energy. Therefore, within the CE, we apply the same
prescription proposed in Ref. \cite{QMC1} to calculate the CE
partition function for medium and heavy isotopes. According to this
prescription, we calculate the LNBCS and LNSCQRPA pairing gaps in
the space spanned by 22 degenerated (proton or neutron)
single-particle levels above the doubly-magic $^{48}$Ca core for Mo
isotopes, whereas the same is done on top of the doubly-magic
$^{132}$Sn core for Dy and Yb nuclei. The obtained partition
function is then combined with those obtained within the
independent-particle model (IPM) by using Eq. (15) of Ref.
\cite{QMC1}, namely
\begin{equation}
{\rm ln}Z'_{\nu} = {\rm ln}Z'_{\nu, tr} + {\rm ln}Z'_{sp} - {\rm
ln}Z'_{sp, tr} ~, \label{IPM}
\end{equation}
where $Z'_{\nu, tr} \equiv Z_{\nu, tr}e^{\beta{\cal E}_0}$ is the
excitation partition function with respect to the ground state
energy ${\cal E}_0$ with $Z_{\nu, tr}$ being the CE partition
function obtained within the LNBCS [Eq. \eqref{ZBCS}] or LNSCQRPA
[Eq. \eqref{ZQRPA}] for 22 degenerated single-particle levels around
the Fermi energy. $Z'_{sp}$ is the CE partition function obtained
within the IPM [See e.g. Eq. (8) of Ref. \cite{QMC1}] for the space
spanned by the levels
from the bottom to $N=$ 126 closed shell, whereas $Z'_{sp, tr}$ is
the same partition function but for the truncated space spanned by 22 levels around
the Fermi energy.

%%%%%%%%%%%%%%%%%%%%%%%%%%%%%%%%%%%%%%
\subsection{\label{Molyb}Results for molybdenum}
%%%%%%%%%%%%%%%%%%%%%%%%%%%%%%%%%%%%%%
\begin{figure}
     \includegraphics[width=11.0cm]{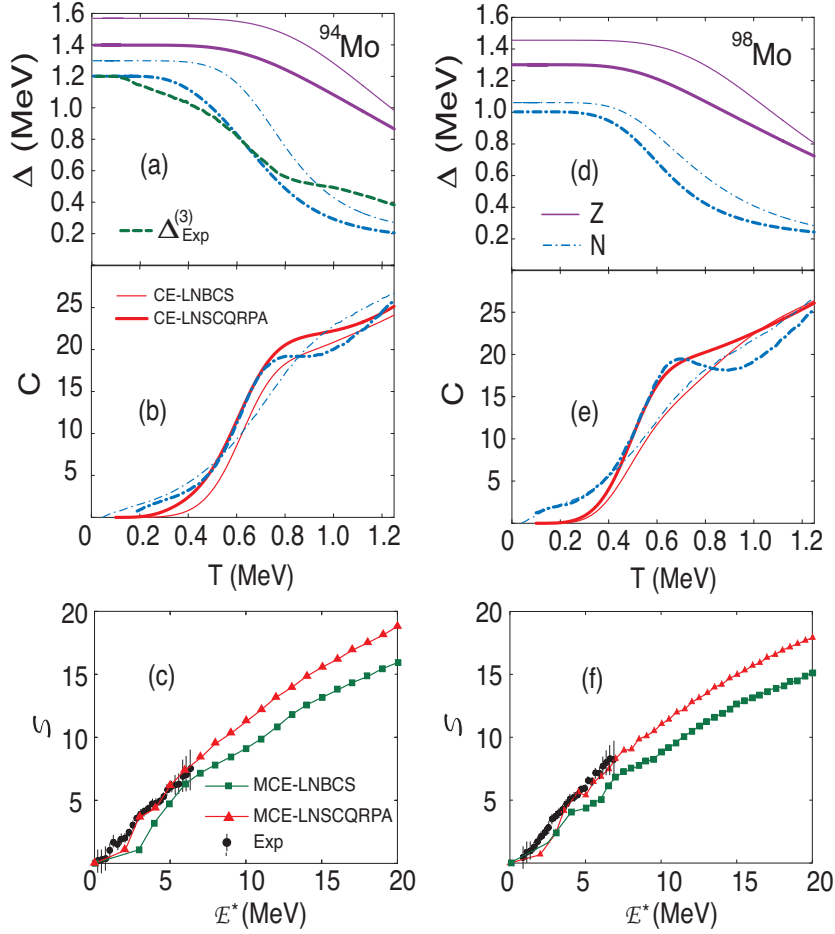}
     \caption{(Color online) Pairing gaps $\Delta$ and heat capacities $C$ obtained within
     the CE as functions of $T$ and entropies ${\cal S}$ obtained within the MCE as functions
     of $E^*$ for $^{94}$Mo (left panels) and $^{98}$Mo (right
     panels). In (a) and (d), the solid and dash-dotted lines denote
     the pairing gaps for protons and neutrons, respectively,
     whereas the thin and thick lines correspond to the CE-LNBCS and
     CE-LNSCQRPA results, respectively. In (b) and (e), the thin and thick
     lines stand for the CE-LNBCS and CE-LNSCQRPA results, whereas the thin
     and thick dash-dotted lines depict the experimental results
     taken from Refs. \cite{Chankova} and \cite{Kaneko},
     respectively. Shown in (c) and (f) are the MCE entropies obtained within
     the MCE-LNBCS (rectangles), MCE-LNSCQRPA (triangles), and
     extracted from experimental data (circles with error bars) of ~Ref.
     \cite{Chankova}.
     \label{Mo}}
\end{figure}
Shown in Fig. \ref{Mo} are the pairing gaps, heat capacities and
entropies for $^{94}$Mo [Figs. \ref{Mo} (a)-\ref{Mo} (c)] and
$^{98}$Mo [Figs. \ref{Mo} (d)-\ref{Mo} (f)] obtained within the
CE(MCE)-LNBCS and CE(MCE)-LNSCQRPA versus the experimental data from
Refs. \cite{Chankova} and \cite{Kaneko}. There is a clear
discrepancy in the heat capacities extracted from the same measured
level density in these two papers [Figs. \ref{Mo} (b) and \ref{Mo}
(e)]. The heat capacity, extracted in Ref. \cite{Kaneko}, clearly
shows a pronounced peak at $T \sim$ 0.7 MeV for both $^{94}$Mo and
$^{98}$Mo, whereas the corresponding quantity, extracted in Ref.
\cite{Chankova}, shows no trace of any peak. The source of the
discrepancy comes from the difference in the scale of excitation
energy $E^{*}$, which was used for extrapolating the measured level
density before evaluating the CE partition function using the
Laplace transformation of the level density. In Ref.
\cite{Chankova}, the level density is extrapolated up to $E^* \sim $
40 - 50 MeV, whereas in Ref. \cite{Kaneko} this is done up to $E^*=$
180 MeV. Given that all the excited states should be included in the
partition function, the energy $E^* \sim $ 40 - 50 MeV used in Ref.
\cite{Chankova} seems to be too low, which might affect the
resulting heat capacity. As Figs. \ref{Mo} (b) and \ref{Mo} (e)
show, the heat capacities predicted by the CE-LNSCQRPA are much
closed to those obtained in Ref. \cite{Kaneko}. They are also
consistent with the FTQMC calculations for other nuclei
\cite{QMC,QMC1}. It is important to emphasize here that quantal and
thermal fluctuations within the CE-LNBCS (LNSCQRPA) indeed smooth
out the SN phase transition. As the result, the pairing gaps [Figs.
\ref{Mo} (a) and (d)] obtained for protons (solid lines) and
neutrons (dash-dotted lines) within both CE-LNBCS (thin lines) and
CE-LNSCQRPA (thick lines) do not collapse at the critical
temperature $T = T_c$ of the SN phase transition, as predicted by
the GCE-BCS, but monotonously decrease with increasing $T$. The
neutron gap in Fig. \ref{Mo} (a) obtained within the CE-LNSCQRPA for
$^{94}$Mo (thick dash-dotted lines) is close to the three-point gap
(dashed lines) obtained in Ref. \cite{Kaneko} by simply
extrapolating the odd-even mass formula to finite temperature. As
has been pointed out in Ref. \cite{Ensemble} such naive
extrapolation contains the admixture with the contribution from
uncorrelated single-particle configurations, which do not contribute
to the pairing correlation. Therefore, to avoid obviously wrong
results at high $T$, such contribution should be removed from the
total energy of the system. Nonetheless, in the low temperature
region ($T<$ 1.3 MeV) as that considered here, where the
contribution of uncorrelated single-particle configurations is
expected to be small, the simple extension of the three-point
odd-even mass formula to $T\neq$ 0 can still serve as a useful
indicator.

%%%%%%%%%%%%%%%%%%%%%%%%%%%%%%%%%%%%%%
\begin{figure}
     \includegraphics[width=9.0cm]{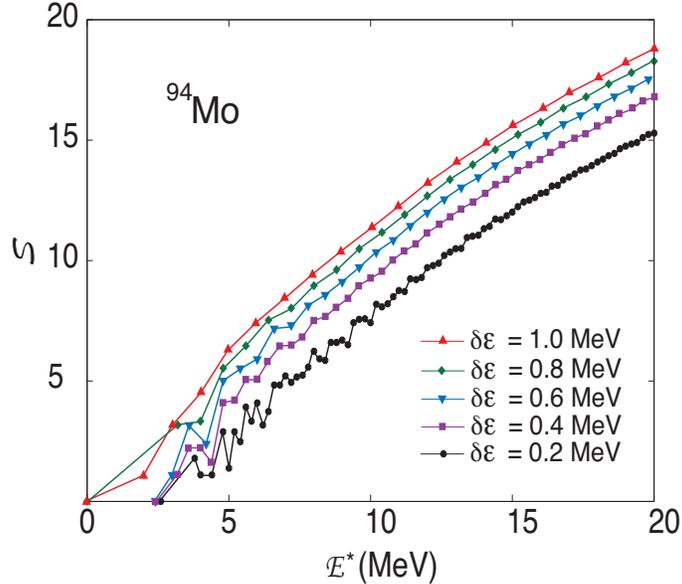}
     \caption{(Color online) Microcanonical entropy as function of
     $E^*$ obtained within the MCE-LNSCQRPA for $^{94}$Mo using various
     values of energy interval $\delta{\cal E}$.
     \label{SMo}}
\end{figure}
%%%%%%%%%%%%%%%%%%%%%%%%%%%%%%%%%%%%%%
As has been discussed in Ref. \cite{CE-BCS}, at low $E^*$ the
genuine thermodynamic observable is the MCE entropy  because it is
calculated directly from the observable level density by using the
Boltzmann's definition (\ref{SMCE}). The experimental MCE entropies
for $^{94, 98}$Mo are plotted in Figs. \ref{Mo} (c) and \ref{Mo} (f)
along with the predictions by the MCE-LNBCS and MCE-LNSCQRPA. These
figures show that the MCE-LNSCQRPA results fit the available
experimental data remarkably well. It is worth mentioning that the
results obtained within the MCE-LNBCS(LNSCQRPA) are sensitive to the
choice of energy interval $\delta{\cal E}$, which is used to
calculate the number of accessible states ${\cal W}({\cal E})$ in
Eq. \eqref{SMCE}. Figure \ref{SMo} shows the entropies obtained
within the CE-LNSCQRPA for $^{94}$Mo using several values of
$\delta{\cal E}$ ranging from 0.2 MeV to 1.0 MeV. It is clear to see
from this Fig. \ref{SMo} that the MCE entropies increase with
increasing $\delta{\cal E}$. In this respect, we found that the
values of $\delta{\cal E}$ = 1 MeV for $^{94}$Mo and 0.7 MeV for
$^{98}$Mo are reasonable to fit the experimental data. The reason
for choosing large values of $\delta{\cal E}$ for these two nuclei
comes from the deficiency of the CE-LNSCQRPA(LNBCS), which includes
only low-lying excited states.

%%%%%%%%%%%%%%%%%%%%%%%%%%%%%%%%%%%%%%
\subsection{\label{DyYb}Results for dysprosium and ytterbium}
%%%%%%%%%%%%%%%%%%%%%%%%%%%%%%%%%%%%%%
\begin{figure}
     \includegraphics[width=11.0cm]{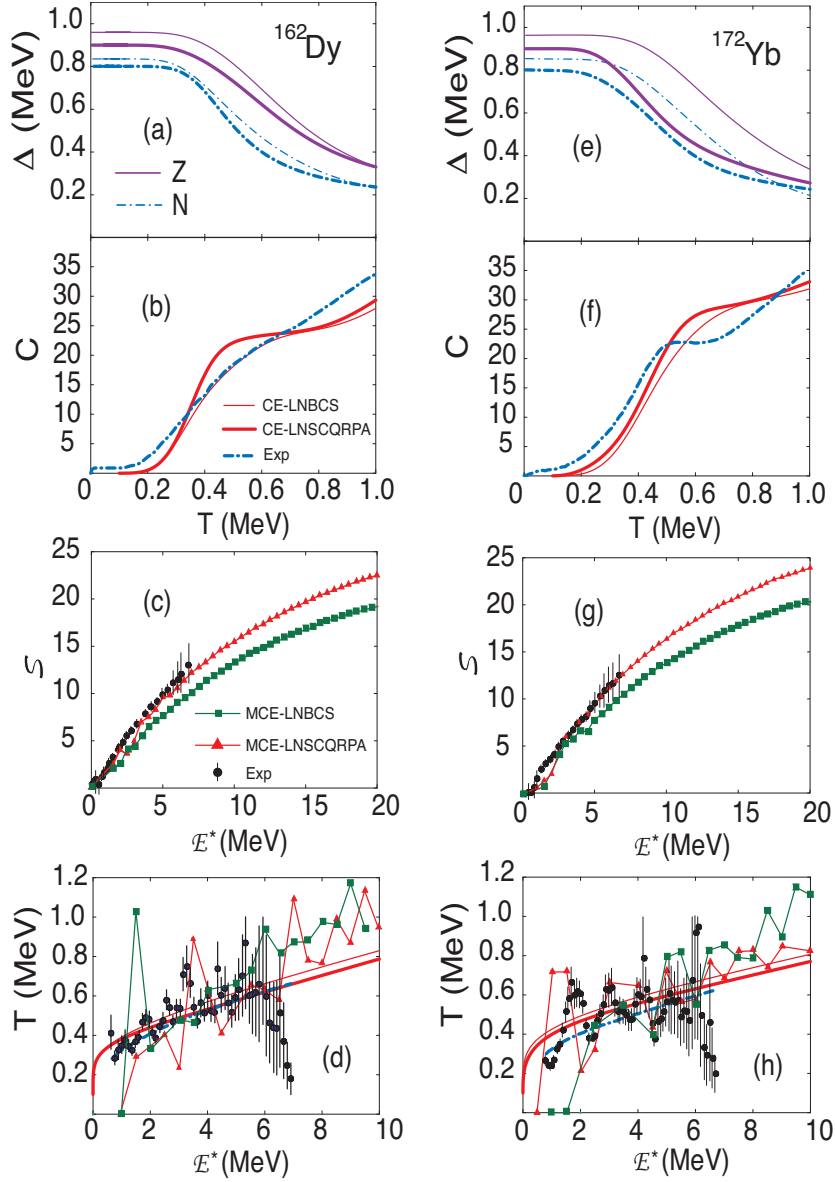}
     \caption{(Color online) (a), (b), (e) and (f): Pairing gaps  $\Delta$, heat capacities
     $C$ as functions of $T$ obtained within the CE;
     (c), (d), (g) and (h): Entropies ${\cal S}$ and
     temperatures $T$ as functions of $E^*$ obtained within the MCE
     for $^{162}$Dy (left panels) and $^{172}$Yb (right panels).
     Notations are the same as those in Fig. \ref{Mo}. Experimental
     data are taken from Ref. \cite{Oslo1}.
     \label{Dy}}
\end{figure}
%%%%%%%%%%%%%%%%%%%%%%%%%%%%%%%%%%%%%%%%%%%%
The results obtained for $^{162}$Dy and $^{172}$Yb are shown in Fig.
\ref{Dy}. Similar to the results for $^{94, 98}$Mo, the CE heat
capacities and MCE entropies obtained within the CE(MCE)-LNSCQRPA
for both $^{162}$Dy and $^{172}$Yb are in good agreement with the
experimental data. The neutron and proton gaps obtained within the
CE-LNBCS (LNSCQRPA) do not collapse at $T=T_c$ but decrease with
increasing $T$ and keep finite at high $T$ even for the two heavy
nuclei considered here. The peak in the experimental heat capacity
near $T=$ 0.4 MeV is seen in $^{172}$Yb, whereas it
disappears in $^{162}$Dy. This is again due to the fact that the
measured level densities for these two nuclei are extrapolated only
up to $E^* = $40 MeV instead of 180 MeV as was done in Ref.
\cite{Kaneko} for other nuclei. This is confirmed by the heat
capacities obtained within the CE-LNSCQRPA (thick solid lines),
which clearly show a peak around $T=$ 0.4 MeV.

In Figs. \ref{Dy} (d) and \ref{Dy} (h), one can see that the MCE
temperatures, extracted from the experimental data (circles with
error bars) by using Eq. \eqref{TMCE},  scatter around the
experimental (thick dash-dotted lines) or theoretical (thick and
thin lines) CE results. The results of calculations with the
MCE-LNBCS (squares) and MCE-LNSCQRPA (triangles)  by using the
same definition (\ref{TMCE}) and $\delta{\cal E}=$ 0.5 also describe
well these values. The results for MCE entropies in Figs. \ref{Mo}
and \ref{Dy} show the importance of the effect beyond the
quasiparticle mean field included in the self-consistent coupling
QRPA vibrations. In fact, the MCE-LNSBCS results for the entropy
clearly underestimate the experimental values. The discrepancy with
the MCE-LNSCQRPA results increases with $E^{*}$ to reach about 20\%
at $E^{*}=$ 20 MeV.
%%%%%%%%%%%%%%%%%%%%%%%%%%%%%%%%%%%%%%
\subsection{\label{NLD} Level density}

\begin{figure}
     \includegraphics[width=13.0cm]{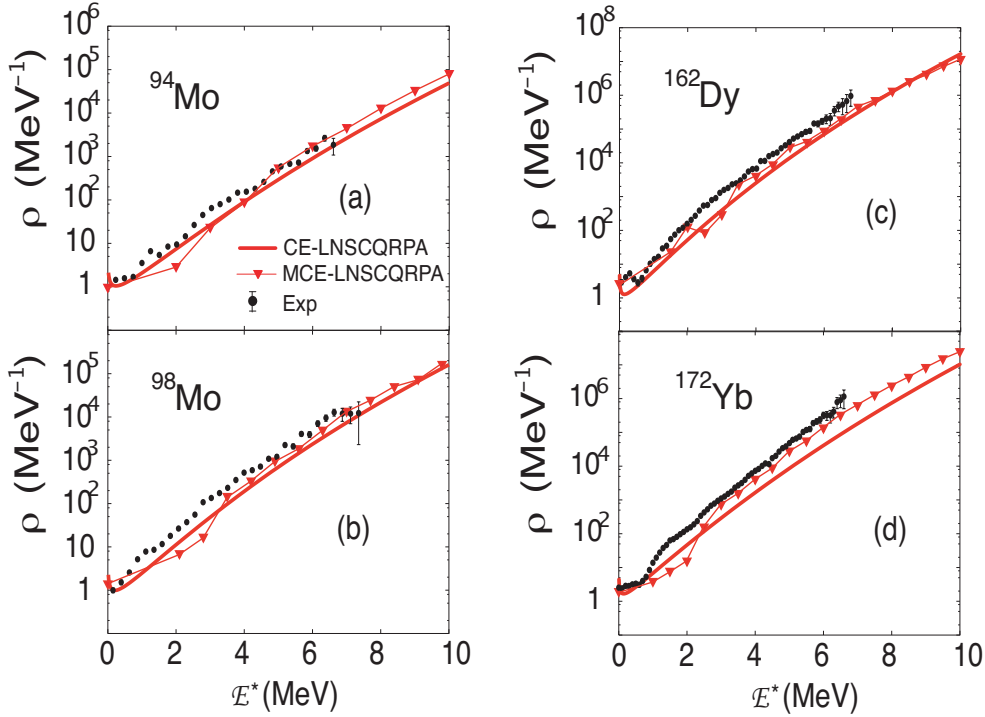}
     \caption{(Color online) Level densities as functions of $E^*$ obtained within the
     CE-LNSCQRPA (solid line) and MCE-LNSCQRPA (triangles) versus the experimental data 
     (circles with error bars) for $^{94}$Mo (a), $^{98}$Mo (b), $^{162}$Dy (c), and $^{172}$Yb (d).
     \label{Rho}}
\end{figure}

The level densities obtained within the CE-LNSCQRPA using Eq.
\eqref{RhoCE} and MCE-LNSCQRPA using Eq. \eqref{RhoMCE} are plotted
in Fig. \ref{Rho} as functions of excitation energy $E^*$ in
comparison with the experimental data~\cite{Oslo1,Chankova}
$\rho_{obs}({\cal E})=\rho_{0}\times{\rm exp}[{\cal S}_{obs}({\cal
E})]$. In the latter $\rho_{0}$ is a normalization factor, which
should be put equal to $1/\delta{\cal E}$ according Eq.
(\ref{Sobs}). However, because of fluctuations in level spacings,
which make the entropy sensitive to $\delta{\cal E}$, the authors of
Ref. \cite{Oslo1,Chankova} chose the values of $\rho_{0}$ to obtain
entropy ${\cal S}_{obs}=$ 0 at $T=$ 0. In this way the value of
$\rho_0$ is set to 1.5 MeV$^{-1}$ for $^{94,98}$Mo \cite{Chankova}
and 3 MeV$^{-1}$ for $^{162}$Dy and $^{172}$Yb \cite{Oslo1}.  Figure
\ref{Rho} shows that the level densities obtained within the
MCE-LNSCQRPA offer the best fit to the experimental data for all
nuclei under consideration. The results obtained within the
CE-LNSCQRPA are closer to the experimental data for $^{94,98}$Mo at
$E^{*}\leq$ 4 MeV, whereas at higher $E^{*}$ the MCE-LNSCQRPA offers
a better performance. The S shape in the MCE-LNSCQRPA level density
at low $E^{*}$ might have come from the fixed value of the energy
interval $\delta{\cal E}$, within which the levels are counted,
according to the definition (\ref{SMCE}), whereas the denominator in
the definition of the CE level density [at the right-hand side of
Eq. (\ref{RhoCE})] depends on $E^{*}$. A larger value $\delta{\cal
E}$ at $E^{*}\leq$ 4 MeV would eventually increase the MCE-LNSCQRPA
level density, improving the agreement with the observed level
density in this region, but there is no physical justification for
doing so. The discrepancy between the CE-LNSCQRPA and experimental
results seems to be larger and increases with $E^{*}$ for $^{162}$Dy
and $^{172}$Yb. This might be due to the absence of the contribution
of higher multipolarities such as dipole, quadrupole etc., which are
not included in the present study and may be important for
rare-earth nuclei. On the other hand, the use of SCQRPA plus angular
momentum~\cite{SCQRPAM}, discussed previously, may also improve the
agreement.
%%%%%%%%%%%%%%%%%%%%%%%%%%%%%%%%%%%%%%
%%%%%%%%%%%%%%%%%%%%%%%%%%%%%%%%%%%%%%%%%%%%%%%%%
\section{\label{Conclu}CONCLUSIONS}

The present article applies the canonical and microcanonical
ensembles of the LNBCS and LNSCQRPA approaches, derived in Ref.
\cite{CE-BCS}, to describe the thermodynamic properties as well as
level densities of several nuclei, namely $^{94, 98}$Mo, $^{162}$Dy
and $^{172}$Yb. The results obtained show that the CE(MCE)-LNSCQRPA
describe quite well the recent experimental level densities and the
thermodynamic quantities extracted for these nuclei by the Oslo
group \cite{Oslo,Oslo1,Chankova,Kaneko}. It confirms that the SN
phase transition is smoothed out in nuclear systems due to the
effects of quantal and thermal fluctuations leading to the
nonvanishing pairing gap at finite temperature even in heavy nuclei
\cite{Moretto,SPA,Zele,MBCS,FTBCS1,Ensemble}. The discrepancy
between the heat capacities obtained within the two different
experimental works, which extrapolate the same experimental level
density to different excitation energies, are also discussed. The
heat capacities obtained within the CE-LNBCS(LNSCQRPA) for all
nuclei show a pronounced peak at $T\sim T_c$, whereas the results
extracted from the same experimental data by Refs. \cite{Chankova}
and \cite{Kaneko} show different behaviors. The better agreement
between the predictions of our approaches as well as those of the
FTQTMC and the results of Ref. \cite{Kaneko} gives a strong
indication to the fact that, to construct an adequate partition
function for a good description of thermodynamic quantities, the
measured level density should be extended up to very high excitation
energy $E^* \sim$ 180 MeV or 200 MeV. The small differences between
the CE(MCE)-LNBCS(LNSCQRPA) results and the experimental data might
be due to the absence of the contribution of higher multipolarities
such as dipole, quadrupole etc., which are not included in the
present study. In order to tackle this issue, the LNSCQRPA plus
angular momentum ~\cite{SCQRPAM} should be used and extended to
included also the multipole residual interactions higher than the
monopole pairing force.  This task remains one of the subjects of
our study in the future.
%%%%%%%%%%%%%%%%%%%%%%%%%%%%%%%%%%%%%%%%%%%%
\begin{acknowledgments}
The numerical calculations were carried out using the FORTRAN IMSL
Library by Visual Numerics on the RIKEN Integrated Cluster of
Clusters (RICC) system. A part of this work was carried out
during the stay of N.Q.H. in RIKEN under the support by the postdoctoral grant
from the Nishina Memorial Foundation and by the Theoretical Nuclear
Physics Laboratory of the RIKEN Nishina Center.
\end{acknowledgments}
%%%%% 2nd revision %%%%%
\appendix
\label{appendix}
\section{MCE results within the Richardson model}
%%%%%%%%%%%
\begin{figure}
     \includegraphics[width=13.0cm]{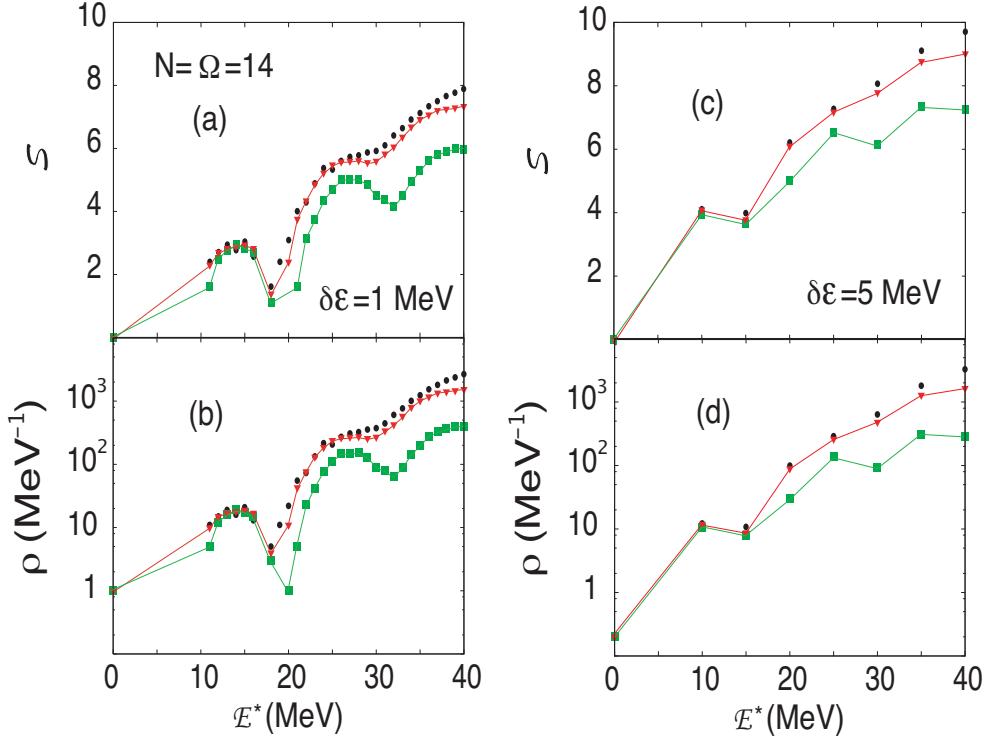}
     \caption{(Color online) MCE entropies and level densities as functions of
     $E^*$ obtained within the MCE-LNBCS (squares), MCE-LNSCQRPA (triangles) 
     versus the exact results for 
     the Richardson model (circles) with $N=\Omega=14$ and $G =$ 1 MeV.
     Results obtained by using the energy bin $\delta{\cal E} =$ 1 MeV 
     are shown in (a) and (b), whereas those obtained by using $\delta{\cal E}
     =$ 5 MeV are shown in (c) and (d). Lines connecting the squares and
     triangles are drawn to guide the eye.
     \label{N14}}
\end{figure}
%%%%%%%%%%%%
The CE-LNBCS and CE-LNSCQRPA has been tested within the Richardson model 
in Ref. \cite{CE-BCS} and the results 
obtained are found in very good agreement with the exact solutions whenever the latter are available.
In order to have more convincing evidences on the accuracy of present approaches, 
we show in Fig. \ref{N14} the MCE entropies and level densities obtained within 
the MCE-LNBCS and MCE-LNSCQRPA versus the exact ones for the Richardson model 
with $N = \Omega$ = 14 and $G$ = 1 MeV. Two 
different values of energy interval $\delta{\cal E}$, 
namely $\delta{\cal E}$ = 1 MeV (left panels) and $\delta{\cal E}$ = 5 MeV (right panels)
are used in calculations. This figure shows that 
the MCE-LNSCQRPA always offers the best fit to the exact results, 
whereas the MCE-LNBCS underestimates the exact ones. 
The decreasing of the entropy as well as level density for the case 
with small value of $\delta{\cal E}$ = 1 MeV shown in Figs. \ref{N14} 
(a) and \ref{N14} (b) is due to the small 
configuration space with $N = \Omega$ = 14 in the present case. This
feature is ultimately related to the problem of using thermodynamics
in very small system with discrete energy levels, where the
temperature may decrease with increasing the excitation energy ${\cal E}^*$
(See Fig. 2 of Ref.~\cite{Ensemble}).
This shortcoming can be effectively overcomed by using a larger 
$\delta{\cal E}$ = 5 MeV. As the result, 
the entropy and level density increase with increasing ${\cal E}^*$ 
as shown in the right panels of Fig. \ref{N14}, although there is no physical 
justification for using such a large value of $\delta{\cal E}$. 

%%%%%%%%%%%%%%%%%%%%%%%%%%%%%%%%%%%%%%%%%%%%%%%%%

%%%%%%%%%%%%%%%%%%%%%%%%%%%%%%%%%%%%%%%%%%%%%%%%%

\begin{thebibliography}{30}
\bibitem{BCS}
J. Bardeen, L. Cooper, and Schrieffer, Phys. Rev.
\textbf{108}, 1175 (1957); M. Sano and S. Yamasaki, Prog. Theor.
Phys. \textbf{29}, 397 (1963).
\bibitem{HFB}
K. Tanabe and K. Sugaware-Tanabe, Phys. Lett. B \textbf{97}, 337
(1980); A.L. Goodman, Nucl. Phys. A \textbf{352}, 30 (1981); K.
Tanabe, K. Sugaware-Tanabe, and H.J. Mang, Nucl. Phys. A
\textbf{357}, 20 (1981); Ibid. \textbf{357}, 45 (1981).
\bibitem{Moretto}
L.G. Moretto, Phys. Lett. B \textbf{40}, 1 (1972); A.L. Goodman,
Phys. Rev. C \textbf{29}, 1887 (1984); J.L. Egido, P. Ring, S.
Iwasaki, and H.J. Mang, Phys. Lett. B {\bf 154}, 1 (1985).
\bibitem{SPA}
R. Rossignoli, P. Ring and N.D. Dang, Phys. Lett. B \textbf{297}, 9
(1992); N.D. Dang, P. Ring and R. Rossignoli, Phys. Rev. C
\textbf{47}, 606 (1993).
\bibitem{Zele}V. Zelevinsky, B.A. Brown, N. Frazier, and M. Horoi,
Phys. Rep. {\bf 276}, 85 (1996).
\bibitem{MBCS}
N. Dinh Dang and V. Zelevinsky, Phys. Rev. C {\bf 64}, 064319
(2001); N. Dinh Dang and A. Arima, Phys. Rev. C {\bf 67}, 014304
(2003); N.D. Dang and A. Arima, Phys. Rev. C {\bf 68}, 014318
(2003); N.D. Dang, Nucl. Phys. A {\bf 784}, 147 (2007).
\bibitem{FTBCS1}
N. Dinh Dang and N. Quang Hung, Phys. Rev. C {\bf 77}, 064315
(2008).
\bibitem{Ensemble}
N.Q. Hung and N.D. Dang, Phys. Rev. C {\bf 79}, 054328 (2009).
\bibitem{Sumaryada}
T. Sumaryada and A. Volya, Phys. Rev. C {\bf 76}, 024319 (2007).
\bibitem{Exact}
R.W. Richardson, Phys. Lett. {\bf 3}, 277 (1963); Ibid. {\bf 14},
325 (1965); A. Volya, B.A. Brown, and V. Zelevinsky, Phys. Lett. B
{\bf 509} (2001) 37.
\bibitem{QMC}
S. Liu and Y. Alhassid, Phys. Rev. Lett {\bf 87}, 022501 (2001);
\bibitem{QMC1}
Y. Alhassid, G. F. Bertsch, and L. Fang, Phys. Rev. C {\bf 68},
044322 (2003).
\bibitem{Dukelsky}
J. Dukelsky, S. Pittel and G. Sierra, Rev. Mod. Phys. {\bf 76}, 643 (2004).
\bibitem{FTPNP}
R. Rossignoli and P. Ring, Ann. Phys. (NY) 235, 350 (1994); R.
Rossignoli, P. Ring, and N. D. Dang, Phys. Lett. B297, 9 (1992); K.
Tanabe and H. Nakada, Phys. Rev. C {\bf 71}, 024314 (2005); H.
Nakada and K. Tanabe, Phys. Rev. C {\bf 74}, 061301(R) (2006).
\bibitem{CSPA}
R. Rossignoli, N. Canosa, and P. Ring, Phys. Rev. Lett {\bf 80}, 1853 (1998).
\bibitem{NPSPA}
K. Kaneko and A. Schiller, Phys. Rev. C {\bf 75}, 044304 (2007); ibib {\bf 76}, 064306 (2007).
\bibitem{MCESPA}
R. Rossignoli, Phys. Rev. C {\bf 54}, 1230 (1996).
\bibitem{Oslo}
E. Melby \emph{et al.}, Phys. Rev. Lett. {\bf 83}, 3150 (1999);  A.
Schiller \emph{et al.}, Phys. Rev. C {\bf 63}, 021306 (R) (2001); E.
Algin \emph{et al.}, Phys. Rev. C {\bf 78}, 054321 (2008).
\bibitem{Oslo1}
M. Guttormsen \emph{et. al}, Phys. Rev. C {\bf 62}, 024306 (2000).
\bibitem{Chankova}
R. Chankova \emph{et al.}, Phys. Rev. C {\bf 73}, 034311 (2006).
\bibitem{Kaneko}
K. Kaneko \emph{et al.}, Phys. Rev. C {\bf 74}, 024325 (2006).
\bibitem{CE-BCS}
N.Q. Hung and N.D. Dang, Phys. Rev. C {\bf 81}, 057302(BR) (2010).
\bibitem{SCQRPA}
N.Q. Hung and N.D. Dang, Phys. Rev. C {\bf 76}, 054302 (2007); Ibid.
{\bf 77}, 029905(E) (2008).
\bibitem{LN}
H. J. Lipkin, Ann. Phys. (NY) \textbf{9} 272 (1960); Y. Nogami,
Phys. Lett. \textbf{15} 4 (1965).
\bibitem{Chemical}
N. Dinh Dang and N. Quang Hung, Phys. Rev. C {\bf 81}, 034301
(2010).
\bibitem{Ericson}T. Ericson, Adv. Phys. {\bf 9}, 425 (1960).
\bibitem{SCQRPAM}
N.Q. Hung and N.D. Dang, Phys. Rev. C {\bf 78}, 064315
(2008).
\bibitem{Bohr}A. Bohr and B.R. Mottelson, {\it Nuclear structure}, Vol.
1 (Benjamin, NY, 1969).
\bibitem{Gilbert}A. Gilbert and A.G.W. Cameron, Can. J. Phys. {\bf
43}, 1446 (1965).
\bibitem{WS}
S. Cwiok {\it et al.}, Comput. Phys. Commun. {\bf 46}, 379 (1987).
\end{thebibliography}
\end{document}